\documentclass[11pt,a4paper]{article} 
\usepackage{jheppub_kim}

\usepackage{multirow, graphicx,amssymb,url,mathrsfs,amsmath}
\usepackage{wrapfig,boxedminipage,setspace,subfigure,epsfig}
\usepackage{amsxtra,amstext,latexsym,dsfont,amsfonts}

             \def\m  {\mu}

\newcommand{\calg}{\mbox{${\cal G}$}} 
 
 \newcommand{\call}{\mbox{${\cal L}$}}
 \newcommand{\caln}{\mbox{${\cal N}$}}

\def\IR{{\hbox{{\rm I}\kern-.2em\hbox{\rm R}}}}
\def\IB{{\hbox{{\rm I}\kern-.2em\hbox{\rm B}}}}
\def\IN{{\hbox{{\rm I}\kern-.2em\hbox{\rm N}}}}
\def\IC{\,\,{\hbox{{\rm I}\kern-.59em\hbox{\bf C}}}}
\def\IZ{{\hbox{{\rm Z}\kern-.4em\hbox{\rm Z}}}}
\def\IP{{\hbox{{\rm I}\kern-.2em\hbox{\rm P}}}}
\def\IH{{\hbox{{\rm I}\kern-.4em\hbox{\rm H}}}}
\def\ID{{\hbox{{\rm I}\kern-.2em\hbox{\rm D}}}}

\def\be{\begin{equation}}
\def\ee{\end{equation}}
\def\ba{\begin{eqnarray}}
\def\ea{\end{eqnarray}}

\def\lra{\Longrightarrow}

\newcommand{\ud}{\mbox{${\mathrm{d}}$}}
\def\det{{\rm det}}

\def\ea{{\it et al}. }

\newcommand{\beq}{\begin{equation}}
\newcommand{\eeq}{\end{equation}}
\newcommand{\bea}{\begin{eqnarray}}
\newcommand{\eea}{\end{eqnarray}}

\newcommand{\tB}{{{B}}}

\newcommand{\tE}{{{ E}}}
\def\dd{{\mathrm{d}}}
\def\r{{\rho}}
\newcommand{\tA}{{{ A}}}
\newcommand{\hJ}{{{\hat{J}}}}

\title{Chiral phase transitions and quantum critical points of the D3/D7(D5) system 
with mutually perpendicular E and B fields at finite temperature and density}

\author[a]{Nick Evans,}
\author[a]{Keun-Young Kim,}
\author[b]{and Jonathan P. Shock}

\emailAdd{evans@soton.ac.uk}
\emailAdd{k.kim@soton.ac.uk}
\emailAdd{jonshock@mppmu.mpg.de}

\affiliation[a]{ School of Physics and Astronomy, University of
Southampton, \\ Southampton, SO17 1BJ, UK.
}
\affiliation[b]{Max-Planck-Institut f\"{u}r Physik (Werner-Heisenberg-Institut), \\
F\"{o}rhringer Ring 6, 80805 M\"{u}nchen, Germany.
}

\abstract{
We study chiral symmetry restoration with increasing temperature and density in gauge theories subject to mutually perpendicular electric and magnetic fields using holography. We determine the chiral symmetry breaking phase structure of the $D3/D7$ and $D3/D5$ systems in the temperature-density-electric field directions. A magnetic field may break the chiral symmetry and an additional electric field induces Ohm and Hall currents as well as restoring the chiral symmetry. At zero temperature the $D3/D5$ system displays a line of holographic BKT phase transitions in the density-electric field plane, while the $D3/D7$ system shows a mean-field phase transition. At intermediate temperatures, the transitions in the density-electric field plane are of first order at low density, transforming to second order at critical points as density rises. At high temperature the transition is only ever first order.
  }


\subheader{SHEP-11-18, $\, \, $ MPP-2011-92}

\begin{document}

\maketitle



\section{Introduction}

There has been considerable interest recently in the chiral symmetry breaking/restoration
phase structure of gauge theories with quarks ~\cite{Erdmenger:2007cm} that can be described 
holographically \cite{Maldacena:1997re,Gubser:1998bc,Witten:1998qj}. 
The simplest model to
study is the 3+1 dimensional ${\cal N}=2$ gauge theory made from adding quark hypermultiplets
to ${\cal N}=4$ super Yang-Mills theory \cite{Grana:2001xn,Bertolini:2001qa}. This theory can be described holographically, in the quenched
limit, by probe $D7$-branes in an AdS$_5\times S^5$ geometry~\cite{Karch:2002sh}. Chiral symmetry breaking can be
induced by the application of a magnetic field which breaks the conformal invariance~\cite{Filev:2007gb}. Temperature, $T$, is introduced by
an AdS-Schwarzschild black hole geometry \cite{Witten:1998qj} and density, $d$,  through a non-zero temporal component of a 
gauge field on the $D7$-brane world-volume dual to a $U(1)$ baryon number background gauge field~\cite{Nakamura:2006xk,Kobayashi:2006sb,Kim:2006gp,Kim:2007zm}. The chiral symmetry 
restoration transition is first order \cite{Albash:2007bk,Erdmenger:2007bn} with temperature and second order with density \cite{Evans:2010iy,Jensen:2010vd}. There
is a critical point linking these behaviours in the $T$-$d$ plane~\cite{Evans:2010iy}. A recent study \cite{Gwak:2011wr} has also looked at the
chiral symmetry breaking phase structure in a background geometry with a running gauge coupling displaying
first order transitions as well as a baryonic phase. These results are potentially
of use in describing 3+1-dimensional gauge theories that break chiral symmetry including, of course, QCD.

The D3/D5 system  with a two dimensional defect is also of interest since it describes a 2+1-dimensional
theory of quark hypermultiplets in the background of 3+1-dimensional ${\cal N}=4$ adjoint matter~\cite{Karch:2000gx,DeWolfe:2001pq,Erdmenger:2002ex,Myers:2008me,Jensen:2010ga,Evans:2010hi}.
This may be relevant to condensed matter systems. Again a background magnetic field
induces chiral symmetry breaking~\cite{Jensen:2010ga,Evans:2010hi}. Here the novel discovery at zero temperature 
but finite density was 
a holographic BKT transition across which the quark condensate grows exponentially with the distance from the 
critical density. This behaviour results from the fact that density and magnetic field are both
dimension 2 in 2+1-dimension and so compete in the deep IR as explained in~\cite{Jensen:2010ga}. Any finite temperature 
converts the transition to a second order mean field transition. Examples of cases
in which a mean field transition interpolates to the BKT point through non-mean field second order transitions
was also explored in~\cite{Evans:2010np}. 

The rich structure of these results is very interesting and it is natural to ask how generic the
phase structures found are. In part to address this question in \cite{Evans:2011mu}  the authors
introduced an electric field parallel to the magnetic field already studied. The electric
field simply adds another parameter acting against the chiral phase (the $E$ field tries
to dissociate the mesons). It is of course also interesting in a condensed matter context
to study the effects of $E$ fields. The chiral phase structure in the $T$-$E$ plane was very similar to
that in the $T$-$d$ plane. The $E$-$d$ structure of the theory was rather singular though with density
not being generated no matter how large a chemical potential was applied. The DBI analysis 
was suspect at the origin of the space where the $U(1)$ baryon number gauge field diverged 
\footnote{In the case of a parallel $E$ and $B$ field there is a Wess-Zumino term in the $D7$ action
which is proportional to $\int \dd^4x E \cdot B$. This is a total derivative and hence if the background $E$ and $B$ fields
fall of sufficiently quickly asymptotically this gives the zero winding number of the $U(1)$ configuration. Presumably
these terms can be neglected as we did in \cite{Evans:2011mu}. Were the fields not to fall off as quickly as $1/r^2$ there might
be additional terms that would effect the analysis. In the case of perpendicular $E$ and $B$ fields 
we study here there are no
such Wess-Zumino terms.}.  

In this paper we wish to add to the investigation of the effects of an $E$ field by studying
the case of an electric field perpendicular to the magnetic field. This case does not show any
signs of singularities in the DBI analysis but is also technically more difficult because of
the presence of a Hall current. It is however possible to extract all the currents analytically~\cite{OBannon:2007in,Kim:2011qh}. 
We will work in the Canonical Ensemble where we treat density as a free parameter.
The crucial transitions on the gravity side are then associated with $D7$ or $D5$-brane embeddings.
If the branes lie flat they describe a chirally symmetric phase. If they bend in
the space transverse to the $D3$ branes they describe chiral symmetry breaking configurations 
\cite{Babington:2003vm}.

Our results are summarized in Fig \ref{3D} where we show the $E$-$T$-$d$ phase space at fixed magnetic field. 
In the D3/D7 case the chiral transitions are first order at low density whilst second order at high density. 
We have identified a critical line between these transition types.
The D3/D5 system is similar except that it has in addition a line of BKT transitions in the $E$-$d$ plane
at zero temperature. The phase boundaries on the $E$-$d$ surface form a square - we show that the BKT
transition is driven by deep IR effects and that they are independent of $E$ until the $E$ field itself
is super-critical when it overcomes the BKT transition entirely, explaining this shape. In this
case there is no interpolation from mean field to a BKT transition.

\section{D-brane setup}

We study a single probe flavour brane embedded in the Poincar\'{e} patch of the Schwarzschild $AdS_5\times S^5$ geometry, generated by a single stack of $N_c\gg 1$ black $D3$-branes. We choose the following metric parametrization:
\begin{eqnarray}
  \dd s^2 = G_{MN} \ud x^M \ud x^N = \frac{r^2}{R^2}\left(-f \dd t^2+\dd \vec{x}^2\right)+ \frac{R^2}{r^2} \frac{\dd r^2}{f}
   + R^2 \dd \Omega_5^2 \ , \quad f(r) = 1-\frac{r^4_H}{r^4}\, ,
\end{eqnarray}
where $R$ is the radius of the $AdS_5$ and $S^5$ spaces, $\dd \vec{x}^2 \equiv \dd x^2 + \dd y^2 + \dd z^2$ and 
$\dd \Omega_5$ is the unit radius $S^5$. 
$r$ goes from the horizon at $r=r_H$ to the boundary at $\infty$. The temperature of the black hole and the associated boundary field theory are both given by:
\begin{equation}
T_H=\frac{r_H}{\pi R^2}\, .
\end{equation}
We will specialize to the cases of $Dq$-probe brane phenomenology with $q=5$ and 7. In the $q=5$ case 
the $D3$ and $D5$ have 2+1 coincident dimensions and the description corresponds to a theory with fundamental matter living on a $2+1$-dimensional defect of the $4D$ CFT and may be used to study the phenomenology of some condensed matter systems~~\cite{Karch:2000gx,DeWolfe:2001pq,Erdmenger:2002ex,Myers:2008me,Jensen:2010ga,Evans:2010hi}. $q=7$, when the $D3$ and $D7$ have 3+1 coincident dimensions, is the canonical model for a flavoured $3+1$-dimensional theory and, with the inclusion of a conformality breaking scale and can be considered as a QCD-like model~\cite{Karch:2002sh,Filev:2007gb,Albash:2007bk,Erdmenger:2007bn,Evans:2010iy,Jensen:2010vd}.

The $Dq$-brane action is given in terms of the metric pulled back onto its worldvolume. This includes the scalar fluctuations of its transverse dimensions and to this we add the gauge field contribution. All fermionic fields will be turned off in what follows. We will also not include the Wess-Zumino term which vanishes for all cases of interest. The DBI action is thus:
\be\label{eq.dbi}
S_{\mathrm{DBI}} = -T_{Dq} N_f \int \dd^{q+1}\xi\, e^{-\phi} \sqrt{-\det(\gamma_{mn})} \, ,
\ee
where $T_{Dq} = (2\pi)^{-q}\alpha'^{-(q+1)/2}$ and $\gamma$ denote the pullback metric plus gauge field
\be\label{eq.gamma}
\gamma_{mn} \equiv \partial_m X^M \partial_n X^N G_{MN}+ 2\pi\alpha' F_{mn}\,. 
\ee

We will consider different choices for the gauge field on the $Dq$-brane, but the one-form gauge potential ($A$, such that  in equation \eqref{eq.gamma} $F=dA$) which captures all phenomena discussed here is~\cite{OBannon:2007in,Kobayashi:2006sb,Filev:2007gb,Erdmenger:2007bn}\footnote{In \cite{Erdmenger:2011hp} an isospin chemical potential was turned on using a pair of $D7$-branes with a $U(2)$ gauge field and the phase structure of this system studied in detail.}
\begin{eqnarray}
  2 \pi \alpha' {A} = \tA_t(\r) \mathrm{d} t  + (- \tE_x t + \tA_x(\r) ) \mathrm{d} x  + (\tB_z x+\tA_y(\r)) \, \dd y \nonumber  \ , \label{AA}
\end{eqnarray}
where the $A_i$, $E_x$, and $B_z$ variables include the factor of $2\pi \alpha'$ and correspond to the background field. $A_t$  introduces finite baryon density, $\tE_x$ is a constant background electric field and $\tB_z$ a magnetic field in the $x$ and $z$ directions respectively. We will drop the $x$ and $z$ indices from now on. $A_x$ and $A_y$ will encode the Ohm and Hall currents generated by the electric field and magnetic field.  This choice of the gauge potential is sufficient to find the vacuum $D7$ and $D5$ configurations with
our $E$ and $B$ field choices (though if we were interested in mesonic states described by fluctuations of the gauge field degrees of freedom we would need a more complete ansatz). In the case of the $D5$-brane, the magnetic field is a scalar, while in the $D7$-brane case it is a pseudovector, pointing in a direction transverse to the electric field. The fact that the electric and magnetic fields are pointing in different directions results in a vanishing Wess-Zumino term. 

To describe the embeddings in a convenient way we parameterize the $S^5$ as 
\begin{equation}
  d\Omega_5^2 = d\theta^2 +\cos^2\theta \ud \Omega_{(q-1)/2}^2 + \sin^2\theta \ud \Omega_{(9-q)/2}^2\, , \label{IntS}
\end{equation}
such that the probe brane wraps the $\frac{q-1}{2}$-sphere and its transverse directions are parametrized by the $\frac{9-q}{2}$-sphere and the $\theta$ direction. 
We will find it useful to make the coordinate transformation
\begin{eqnarray}
  \frac{\dd r^2}{r^2f} \equiv \frac{\dd w^2}{w^2}
   \ \lra \  \sqrt{2} w = \sqrt{r^2 + \sqrt{r^4 - r_H^4}}\ , \label{rtow}
\end{eqnarray}
with $\sqrt{2} w_H = r_H$. This change makes the presence of a flat 6-plane
perpendicular to the horizon manifest. We will then write the
coordinates in that plane as $\rho$ and $L$ according to
\begin{eqnarray}
  w = \sqrt{\rho^2 + L^2}\ ,  \quad \rho \equiv w\cos\theta \ ,
  \quad L \equiv w\sin\theta \ ,
\end{eqnarray}
The metric is then
\begin{eqnarray}
   \dd s^2 = \frac{w^2}{R^2}(- \bar{g}_{tt} \dd t^2 + \bar{g}_{xx} \dd \vec{x}^2) 
         + \frac{R^2}{w^2} (\dd \rho^2 + \rho^2 \dd \Omega_3^2
         + \dd L^2 + L^2 \dd\Omega_1^2) \ ,
\end{eqnarray}
where
\begin{eqnarray}
\bar{g}_{tt} = \frac{(w^4 - w_H^4)^2}{ w^4 (w^4+w_H^4)}\ ,  \qquad
\bar{g}_{xx} = \frac{w^4 + w_H^4}{  w^4} \ .
\end{eqnarray}

In this coordinate system, we study the embedding scalar $L=L(\rho)$ 
assuming all other scalars to be constant (and indeed zero by symmetry).
Therefore, the DBI action at zeroth order in fluctuations \eqref{eq.dbi} effectively describes the dynamics of a single scalar field plus a gauge field as a function of a single radial direction
\begin{equation}
  S_{\mathrm{DBI}} =  \caln \int \ud t \ud \vec{x} \int \ud \rho \, 
  \call\left[L(\rho),L'(\rho), \tA'_{\mu};\rho\right]\, ,
\end{equation}
where $\caln = N_f T_{Dq} g_s^{-1} \int \ud \Omega_{(q-1)/2}$.

\subsection{The Currents}

We will begin by determining the currents induced by the background $E$ and $B$ fields. 
Since $A_\mu$ occurs only through its derivative we define the associated conserved current as
\begin{equation}
  \hJ^\m = \frac{\partial{\call}}{\partial{\tA'_\mu}} \ ,
\end{equation}
and from this we can define the Legendre-transformed Lagrangian 
\begin{equation}
  \call_{\mathrm{LT}} \equiv \call - \hJ^\mu \tA'_\mu =  - \frac{\mathrm{sgn}(\xi)}{\sqrt{-G_{tt}} G_{xx}}\sqrt{\xi \chi - a^2}\sqrt{G_{\rho\rho}+G_{LL}L'^2}\, , \label{LT}
\end{equation}
where $d \equiv \hJ_t$ and 
\begin{align}
  \xi &= -G_{tt} G_{xx}^2 - G_{tt}\tB^2 - G_{xx} \tE^2 \,,\\ 
  \chi &= - d^2 G_{tt} - G_{tt} G_{xx}^{\frac{q-1}{2}} G_{\Omega\Omega}^{\frac{q-1}{2}} - G_{xx}(\hJ_x^2 + \hJ_y^2) \,, \\
  a &= -d G_{tt} \tB +   G_{xx} \hJ_y \tE  \, .
\end{align}

In order to fix the values of the currents for a given external field configuration we have to distinguish between the finite and zero temperature configurations.

\subsubsection{Finite temperature}

With finite electric field and at finite temperature there is always a position, $w_s > 0$, where $\xi$ changes its sign, which we call the {\it singular shell}, \cite{Erdmenger:2007bn}. In order to keep the Lagrangian real 
we have to require that both $\chi$ and $a$ change their sign at $r_s$. i.e. 
$\chi(w_s) = a(w_s) =0$. The logic is that $a$ must vanish at the point where $\xi$ vanishes to keep the action real
since $a^2$ enters with a negative sign. Further, so that the action is real in the regime where $\xi$ is negative $\chi$ must also have switched sign so that $\xi \chi$ is positive.
This constraint of reality is enough to determine the currents completely from the background field values \cite{OBannon:2007in}
\begin{align}
  \hJ_x & = \frac{\sqrt{2 d^2 (\calg+T^4) + \frac{4(\calg+T^4)(\calg+T^4+2\tB^2) \rho_s^{(p-1)}}{\calg + \sqrt{\calg^2 - T^8}}}}{\calg + T^4 + 2 \tB^2} \tE \,, 
  \label{Jx} \\
  \hJ_y & = - \frac{2 d \tB}{\calg+T^4+2\tB^2} \tE \,, \label{Jy}
\end{align}
where  $T=\pi R T_{H}$, $\rho_s$ is the position at which the embedding touches the singular shell, and
\begin{equation}
  \calg \equiv -\tB^2 + \tE^2 + \sqrt{-4\tB^2 \tE^2 + (T^4 + \tB^2 + \tE^2)^2} \, .
\end{equation}

We see therefore that when we have finite magnetic field, electric field and baryon density at finite temperature there is both an Ohm current, $J_x$, and a Hall current, $J_y$, caused by the Ohm current moving transverse to the magnetic field. 
The Ohm current has two contributions - one is proportional to $d$ and
represents the quark density in the vacuum reacting to the electric field. The second part of the current is due to the pair-creation of quarks and anti-quarks from the vacuum. Note that the Hall current only has a non-zero contribution
from the quark density. This just reflects the fact that the Hall currents for the pair created positive and negative charges cancel out completely. 

\subsubsection{Zero temperature}

Unlike at finite temperature \cite{OBannon:2007in,Kim:2011qh}, at zero temperature there are two cases which have to be studied separately:

$\mathbf{\tE \ge \tB}$: When the electric field is larger than the magnetic field there is a singular shell at finite radius, defined via $\xi=0$, just as in the finite temperature case. So, from $\chi=a=0$, 
\begin{equation}
  \hJ_x = \sqrt{d^2\left(1-\frac{\tB^2}{\tE^2}\right) 
       +\left(\tE\sqrt{1-\frac{\tB^2}{\tE^2}}\right)^{(q-1)/2}}  \,, \quad 
  \hJ_y = -\frac{d\tB}{\tE}\, , \label{Hall0}
\end{equation}
where we have used $\rho_s^2 = \sqrt{\tE^2-\tB^2}$, since the embedding is flat in this case.

$\mathbf{\tE \le \tB}:$ When the electric field is smaller than the magnetic field, the equation $\xi=0$ does not yield a finite radius singular shell ($r_s = 0$), so the reality condition is not $\xi=a=0$ at $\r=0$, 
but $\xi\chi \ge a^2$ for all $\rho$. 
\begin{align}
  \xi\chi - a^2 &\sim 
  ((\rho^2+L^2)^2+\tB^2-\tE^2)(\rho^4+d^2 - \hJ_x^2 -\hJ_y^2) - (d\tB + \hJ_y\tE)^2 \\
  & \ge -\hJ_x^2(\tB^2 - \tE^2) - (\hJ_y \tB + d \tE)^2 \, .
\end{align}
Thus the condition $\xi\chi - a^2 \ge 0$ implies that
\begin{equation}
  \hJ_x = 0\,, \qquad \hJ_y = - \frac{d \tE}{\tB} \,. \label{Hall1}
\end{equation}
Note that $\xi=a=0$ at $\rho_s=0$ does not yield a consistent result.
This result holds for both the $D7$ and $D5$-branes. 
That $J_x$ here is zero but the Hall current is non-zero is a standard steady state Hall
effect result at large magnetic field. 
$J_x = 0$ because the Lorentz force due to B and the electric force due to E at large $\tB$ are equal and opposite. 
However, if $\tE>\tB$, the force due to $\tE$ starts dominating over the Lorentz force and a finite Ohm current is generated \eqref{Hall0}. 

If we consider $\hJ_x$ as an ``order parameter'' of the quantum phase transition 
(nonzero at $\tE > \tB$ and zero otherwise) then there is a second order phase transition at $\tE=\tB$ whose critical exponents are, from \eqref{Hall0}, 
\begin{align}
  \hJ_x &\sim (\tE-\tB)^{1/2} \,, \quad (D5)  \label{O1}\, ,\\
  \hJ_x &\sim (\tE-\tB)^{1/2} \quad (D7, d \ne 0)\,, \label{O2} \\ 
  \hJ_x &\sim (\tE-\tB)^{3/4}  \quad (D7, d=0)\,. \label{O3}
\end{align}
Without density, the critical exponents are understood on dimensional ground ($1/2$ for $D5$ and $3/4$ for $D7$). At finite density, the phase transition is governed by a density effect ($1/2$ for both $D5$ and $D7$). For the $D5$-brane case, they happen to be the same. 
Note that a thermal insulator-conductor transition is of first order 
as shown in section \ref{sec3D}\footnote{Note that the zero-temperature insulator conductor transition can also be first order if we include finite mass for the fundamental matter. In \cite{Mas:2009wf} it was shown that at zero temperature and zero magnetic field, the transition from Minkowski, to singular shell embeddings has a region of first order behaviour as well as second order behaviour.}
In the following sections, we will show that these critical exponents 
remain also valid for the chiral condensate.

\subsection{The embedding and chiral condensate}

Having found the currents it now only remains for us to find the $D7$ or $D5$-brane embedding in the geometry.
The $U(1)$ chiral (R-) symmetry of the gauge theory is dual to the $U(1)$ rotational symmetry 
in the $\Omega_{(9-q)/2}$ of \eqref{IntS}. 
This rotational symmetry is broken 
when the embedding is not $L=0$.  
The equation of motion of $L(\rho)$ is obtained by varying \eqref{LT} after plugging in \eqref{Jx} and \eqref{Jy} (or \eqref{Hall0}, \eqref{Hall1}).  
In general the classical embedding behaves at large $\rho$ (UV) with two parameters 
$m,c$
\begin{equation}
  L_{\rho\rightarrow \infty}  \sim m + \frac{c}{\rho^{(q-3)/2}} \, .
\end{equation}
We are most interested in studying the phenomenon of spontaneous chiral symmetry breaking and thus require the symmetry to be restored in the UV. For this we must impose that $m=0$ and determine the value of $c$ by demanding IR regularity of the embedding solution. If the regular (and lowest free energy solution) gives $c=0$ then the system is in a chirally symmetric phase, whereas for $c\ne 0$ the chiral symmetry has been spontaneously broken. At zero temperature
and density we impose the IR boundary condition $L'(0)=0$ shooting out to find regular solutions with $m=0$.
At zero temperature but non-zero density we seek solutions that end in the IR at the origin so
$L(0)=0$. Finally at non-zero temperature we shoot out from points 
on the singular shell and seek solutions that tend to $L=0$ at large $\rho$. Points
where there are phase transitions can be identified by plotting the parameter $c$ as the background fields are varied.
For more technical details we refer to \cite{Evans:2010iy,Evans:2010hi,Evans:2011mu}.

\section{The $D3/D5$ and $D3/D7$ phase diagrams}
Here we study the phase diagram of the above $D5$ and $D7$-brane systems in the $d$, $\tE$, $\tB$, $T$ directions. The two brane systems will be discussed in parallel, even though we will see that there are some stark differences between the phase structures. We will always be studying the cases of massless fundamental matter and thus there are four dimensionful parameters. Because the underlying theory is conformal, we can use one of these parameters to set the scale for all other parameters. We chose $\tB=1$ and set all other scales by this value. Rather than giving the full three-dimensional phase diagram, to start with we will focus on two dimensional sections of it, contrasting the different behaviours in the two brane systems.

\subsection{$T$-$d$ phase diagram at finite magnetic field}\label{sec:TD}

\begin{figure}[]
\centering
 \subfigure[D3/D5]
  {\includegraphics[width=6cm]{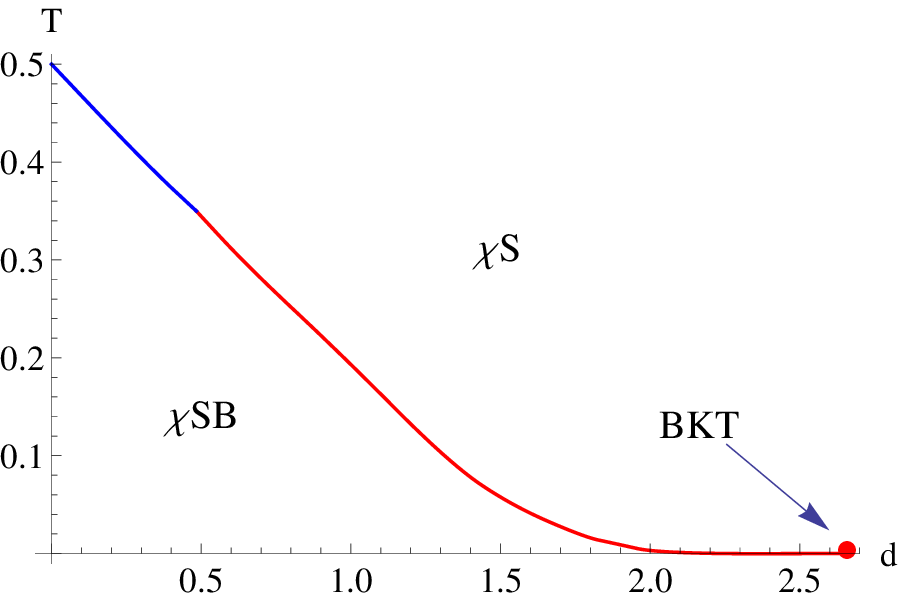}}
  \subfigure[D3/D7] 
   {\includegraphics[width=6cm]{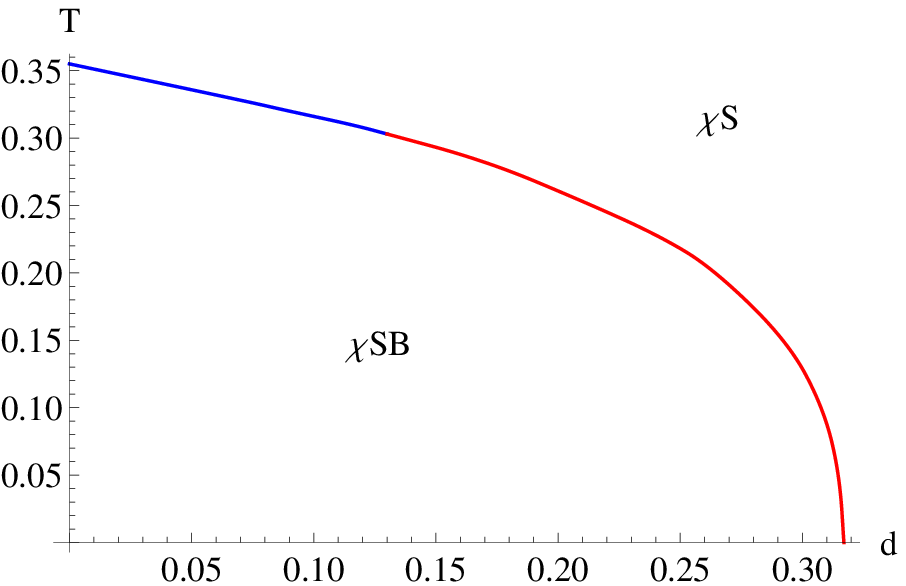}}
  \caption{$T$-$d$ phase diagrams at $\tB=1$ for the $D5$ (a) and $D7$ (b) probes. The red lines correspond to second order phase transitions and the blue to first order phase transitions. There is an isolated BKT phase transition in the $D5$-brane case at exactly $T=0$ due to the competition between $\tB$ and $d$ which in $2+1$ dimensions have the same scaling dimension. The BKT phase transition is absent in the $D7$ case because the dimensions are $[d]=3$ and $[\tB]=2$ which means that such a competition cannot arise and thus the IR phenomenology is dominated by $d$.} \label{Td}
\end{figure}

The phase diagram at finite chemical potential, temperature and magnetic field has been studied in detail in \cite{Evans:2010iy,Evans:2010hi,Jensen:2010vd,Jensen:2010ga}. In the following we focus on the canonical ensemble, the phase diagram of which is shown in Fig \ref{Td}.

The finite magnetic field increases the binding energy of the fundamental charges and thus stabilizes the mesons, whereas the thermal fluctuations at finite temperature act to destabilize them. The finite charge density means that one can never have fully stable bound states and thus the width of all mesonic fluctuations is finite. The transition is thus not a meson melting phase transition (as is the case at zero baryon density) but a chiral symmetry breaking phase transition. For large baryon densities or temperatures, the embedding is always the trivial embedding but the transition from the trivial to the non-trivial embedding may be first order or second order in the $D7$-brane case, with the addition of a BKT phase transition at $T=0$ in the $D5$ brane case\footnote{ At high temperature 
the first order phase transition line turns out to be unstable 
thermodynamically, which can be addressed 
in the grand canonical ensemble more properly~\cite{Evans:2010iy}. 
However, here we deal only with the canonical ensemble.
}. This has already been discussed extensively in~\cite{Jensen:2010ga,Evans:2010hi,Evans:2010np}\footnote{A BKT transition in the back-reacted $U(2)$ Einstein-Yang-Mills theory in 4+1 dimensions has been discussed in \cite{Erdmenger:2011hp}.}.

\subsection{The phase diagram with finite $E$  and $B$}

The inclusion of both finite electric and magnetic field has been discussed previously in  \cite{Evans:2011mu} however in that case the electric and magnetic fields were parallel and thus there was no Hall current. In the current paper we study the case where the two are perpendicular, giving a very different phenomenology and a somewhat different phase structure. In the following sections we will deal with various slices in the finite $E$ part of the phase diagram.

The phenomenology of branes with finite electric fields has been covered extensively in \cite{Karch:2007pd,Erdmenger:2007bn,Albash:2007bq} and in \cite{Kim:2011qh} the interpretation of an open-string horizon with a membrane interpretation was elucidated.

\subsubsection{Transition on the $E$-axis: $T= d = 0$}

In section \ref{sec:TD} we discussed the $T$-$d$ phase diagram. By looking at the behaviour on the $E$-axis (again, at finite, fixed $\tB$), that is at $T=d=0$ we can get an idea of the full $3D$ phase structure. Taking the $T=d=0$ case of the Legendre transformed Lagrangian,\eqref{LT}, and specializing to the $E$-axis we get the simplified Lagrangian:
\begin{eqnarray} \label{zeroT}
  L_{\mathrm{LT}} =
        \rho^{(q-1)/2}\sqrt{\left(1+\frac{1- \tE^2}{ (\rho^2+L^2)^2}\right)}\sqrt{(1+ L'^2)} \ ,  
\end{eqnarray}
where we are restricting to the case where $\tE<1$ (i.e. $\tE < \tB$) such that there is no singular shell and no current induced \eqref{Hall1}. For $\tE>1$ the embedding is always flat and thus there is no condensate whereas for $\tE<1$ there is a finite condensate which vanishes as $\tE\rightarrow 1$. 

\begin{figure}[]
\centering
 \subfigure[Embeddings with $m = 0$. 
    Colors match large dots of (b). ]
   {\includegraphics[width=5cm]{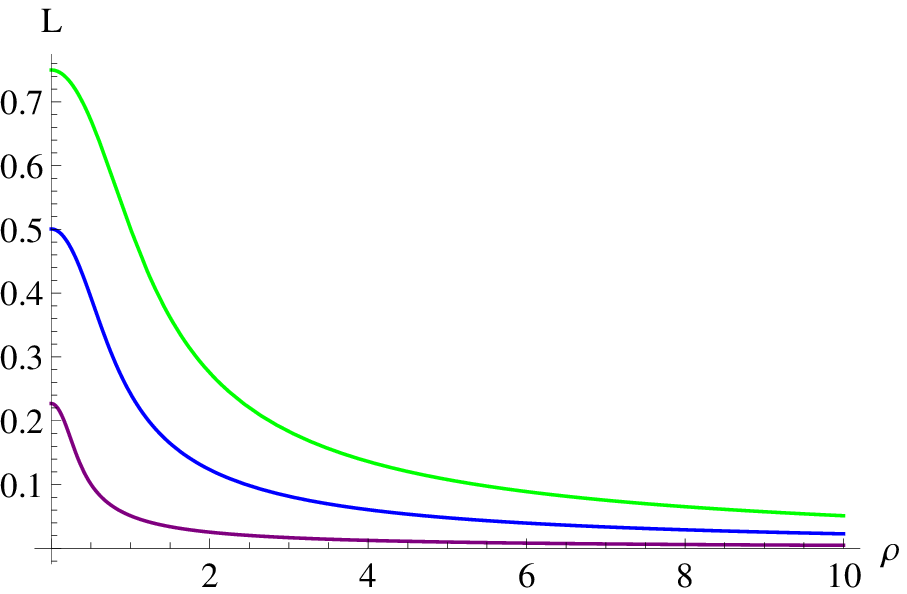}}
  \subfigure[ $c$ vs $E$. Red dots are numerical 
  values and the back curve is $ c = 0.58(1- E^2)^{1/2}$] 
   {\includegraphics[width=5cm]{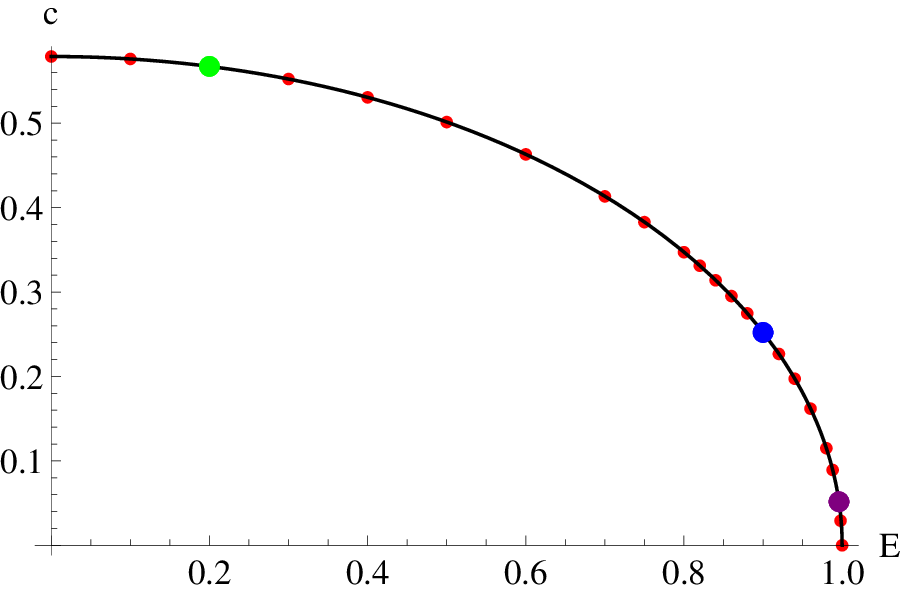}}
  \caption{Second order phase transition for the $D3/D5$ system 
  with $B=1$ varying $E$ at $T=d=0$.} \label{cED5}
\end{figure}
\begin{figure}[]
\centering
 \subfigure[Embeddings with $m = 0$.
  Colors match large dots of (b). ]
  {\includegraphics[width=5cm]{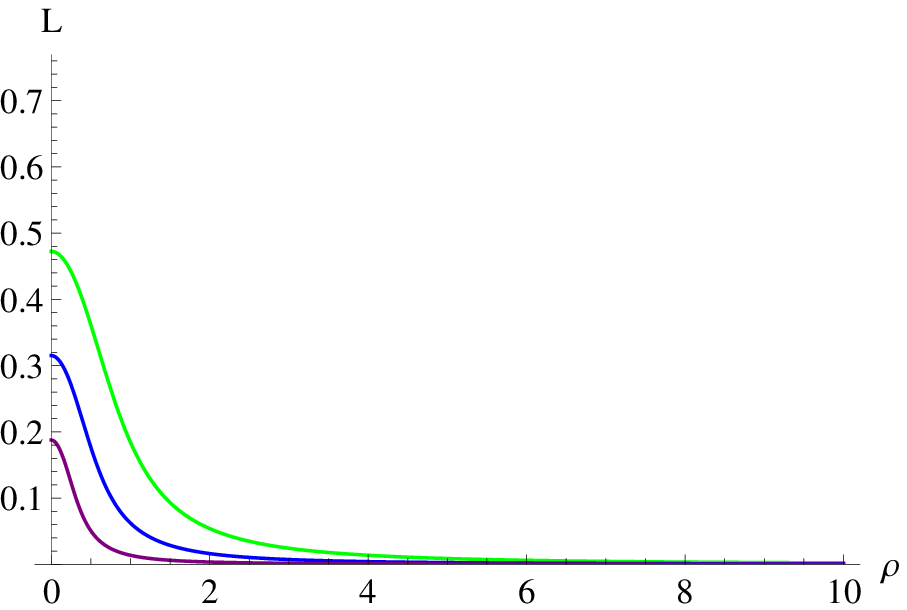}}
  \subfigure[ $c$ vs $ E$. Red dots are numerical 
  values and the back curve is  $ c = 0.225 (1- E^2)^{3/4}$] 
   {\includegraphics[width=5cm]{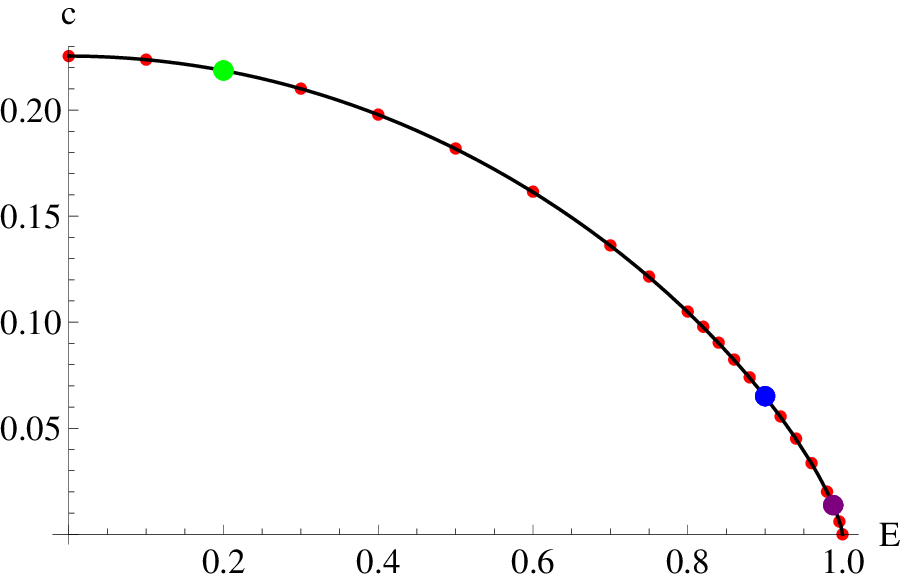}}
  \caption{Second order phase transition for the $D3/D7$ system
  with $B=1$ varying $E$ at $T=d=0$.} \label{cED7}
\end{figure}

In this case the critical exponent of the system can be computed analytically. 
If we substitute $1-\tE^2$ for $\mathbb{B}$, then the 
actions are identical to the case at finite $\mathbb{B}$ and $T=\tE=0$ formally. 
This is nothing but a Lorentz transformation. 
On dimensional grounds the condensate must be related to the electric field as:
\begin{eqnarray}
  &&c \sim (\tE_c - \tE)^{(q-1)/8} \, .
\end{eqnarray}
The proportionality constant is determined by numerical analysis at any one point of $\mathbb{B}$. They turn out to be
\begin{eqnarray}
  && c = 0.580 (1- \tE^2)^{1/2}  \qquad (D5)\ ,  \label{D5c} \label{cE5}\\
  && c = 0.225 (1- \tE^2)^{3/4}  \qquad (D7)\, ,  \label{D7c} \label{cE7}
\end{eqnarray}
where $\tE_c = 1$.
The $D3/D5$ system has the mean-field exponent $1/2$, while that of the $D3/ D7$ shows the non-mean-field exponent 3/4. This is in agreement with the Ohm currents \eqref{O1} and
\eqref{O3} which are also understood on dimensional grounds.

We are able to check our numerical procedures against the analytic results and this is shown in Fig \ref{cED5} and Fig \ref{cED7}. In each of these plots, $(a)$ shows the embedding configuration for a number of different values of the electric field strength, $\tE$, while $(b)$ gives a plot of the condensate as a function of $\tE$ going from 0 to 1. The large coloured dots in the $(b)$ plots correspond to the three embedding configurations in $(a)$. The black curves (\eqref{cE5} and \eqref{cE7}) agree with the red dots, which are numerical values obtained for various $\tE$.   

\subsubsection{$T-E$ plane}

\begin{figure}[]
\centering
\subfigure[D3/D5. the critical exponent at T=0 is $1/2$]
  {\includegraphics[width=6cm]{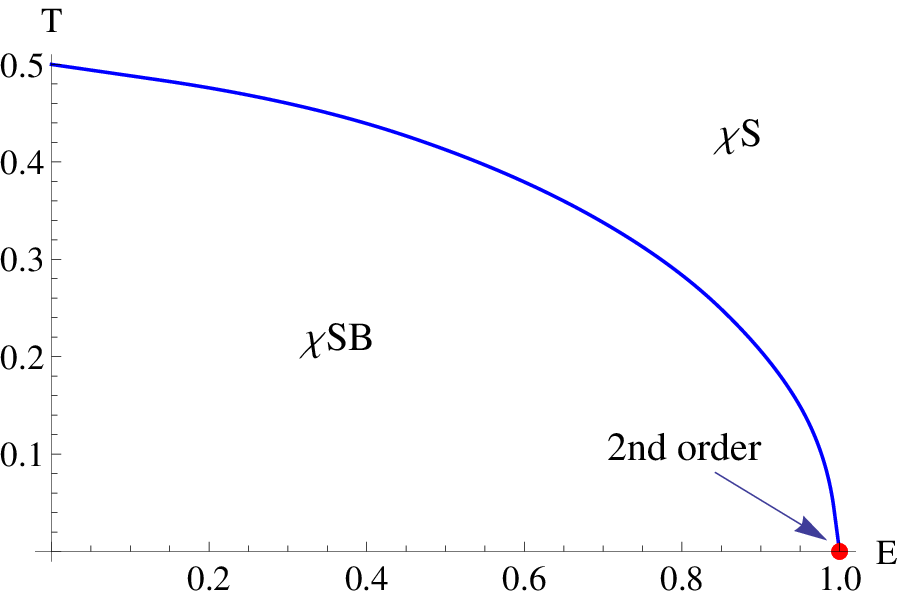}}
\subfigure[D3/D7. the critical exponent at T=0 is $3/4$]
  {\includegraphics[width=6cm]{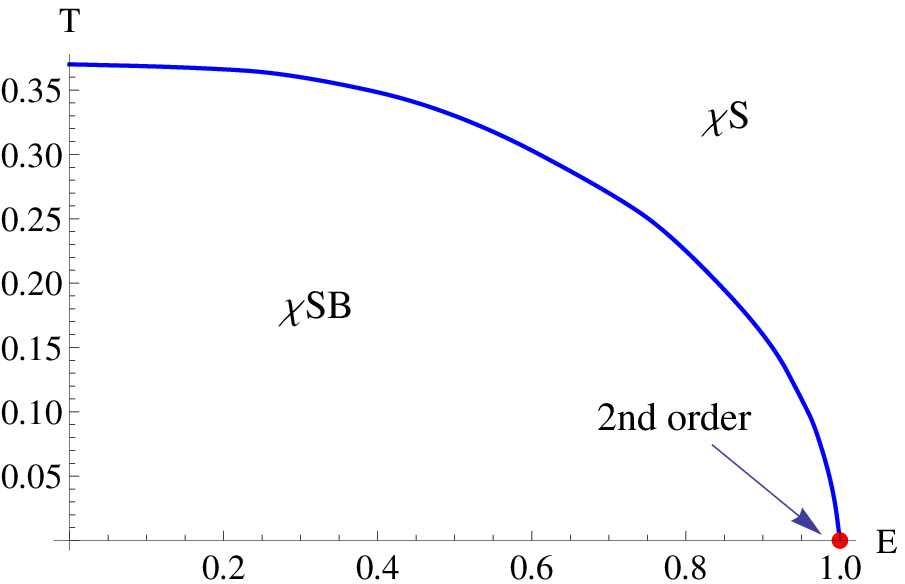}}
  \caption{$T$-$E$ phase diagram at $B=1$ and $d=0$. All fist order transitions(blue) except one second order point(red) at $T=0$} \label{TE}
\end{figure}

The Lagrangian \eqref{LT} with $d=0$ is rather complicated and not terribly illuminating so we don't present it in full here. Both  electric field and finite temperature act to destabilize the fluctuations of the fundamental bilinear operator. The critical line in the $T-E$ plane, shown in Fig \ref{TE}, corresponds to a first order phase transition between a chirally symmetric phase and a chiral symmetry broken phase. This first order transition becomes a second order transition at T=0 which is of mean field type in the $D5$ case and a non-mean field transition (with critical exponent $3/4$) in the $D7$ case.

\subsubsection{$d-E$ plane}
At zero temperature the Lagrangian \eqref{LT} simplifies greatly, with the currents given in \eqref{Hall0}.
\begin{equation}
  \call_{LT} = \sqrt{d^2\left(1-\frac{\tE^2}{\tB^2}\right) 
  + \rho^{q-1}\left(1+\frac{\tB^2-\tE^2}{(\rho^2 + L^2)^2}\right)}
  \sqrt{1+L'^2}\, .
\end{equation}
Although the Lagrangian is simplified, the numerics at $T=0$ are somewhat more complicated than the finite temperature case. At $T=0$ we are not able to integrate the equations of motion from the IR at $r=0$ and instead have to solve from the UV. At finite temperature the singular nature of the equation of motion at the horizon means that given a single boundary condition plus the condition of regularity we can shoot out to the UV and pick off the solution which has zero mass. Shooting from the UV into the IR however we choose the zero mass solution and then have to find the value of the condensate such that the brane embedding hits the point $r=0$. Because of the singular nature of the equation of motion at this point, the numerics are somewhat sensitive to the accuracy which we require. The second subtlety is that because there are actually an infinity of solutions which hit the singularity (associated with the self-similar spiral structure in the $c-m$ plane, \cite{Filev:2007gb}) we must be careful that we pick the one associated with the lowest free energy.

\begin{figure}[]
\centering
\subfigure[D3/D5. The dotted line is a BKT transition and 
the solid line is mean-field 2nd order transition.]
  {\includegraphics[width=6cm]{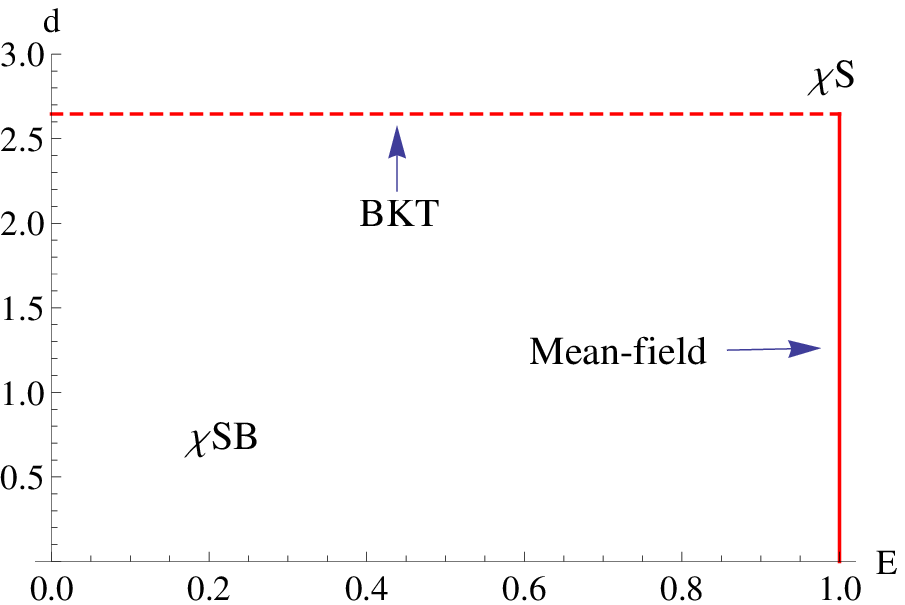}}
\subfigure[D3/D7. All mean field 2nd order except one point ($d=0,E=1$).]  
  {\includegraphics[width=6cm]{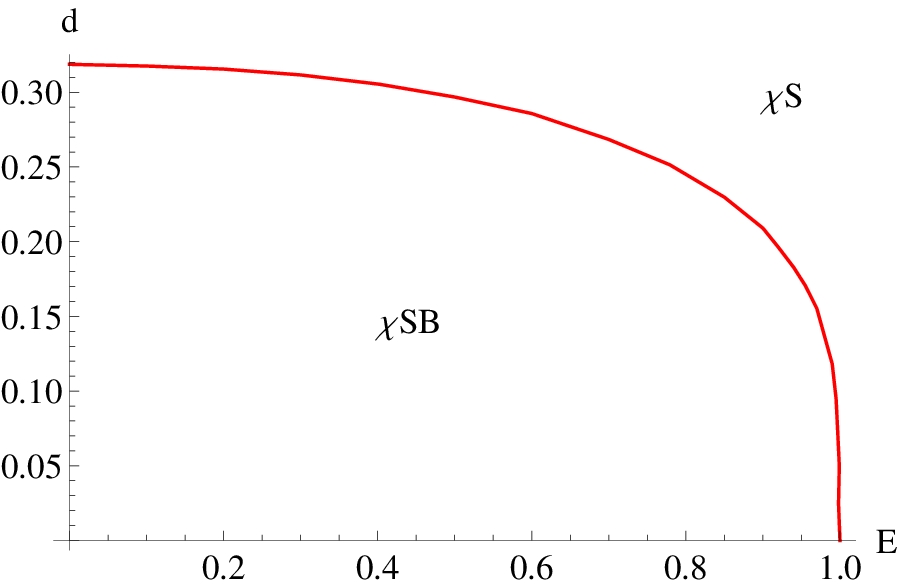}}
  \caption{The $d$-$E$ phase diagram for the $D5$ (left) and $D7$ (right) systems. Chiral symmetry is broken inside the enclosed region.
  $B=1$ and $T=0$.} \label{Ed}
\end{figure}

\begin{figure}[]
\centering
\subfigure[$c(E,d)$. BKT transitions on the line $d = \sqrt{7}$ for 
  ($ 0 \le E < 1$). Mean field second order transition on the line 
  $E=1$ for ($0<d<\sqrt{7}$). ] 
  {\includegraphics[width=6cm]{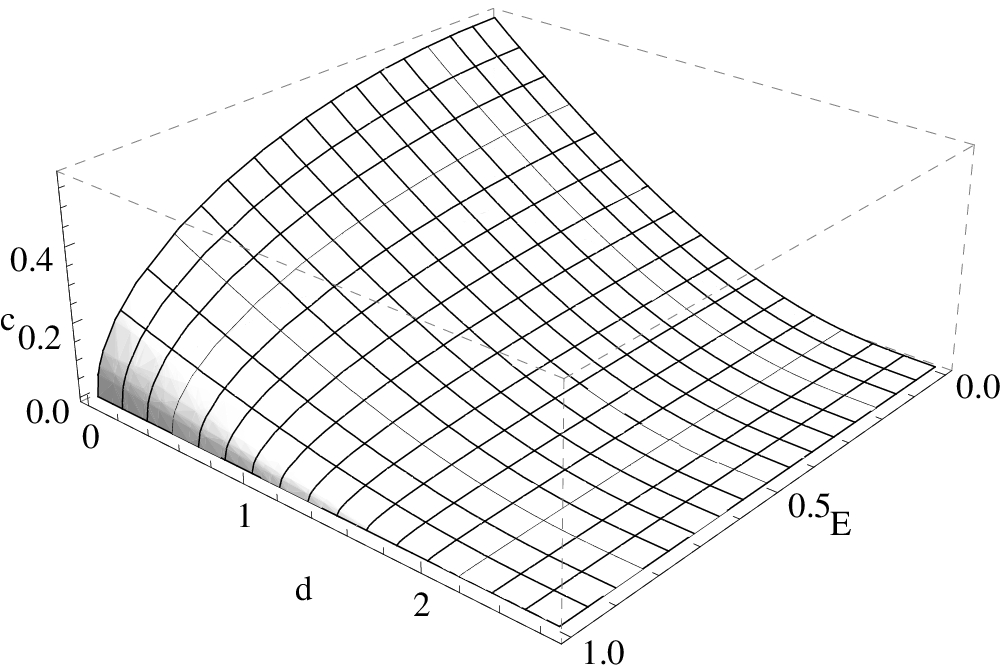}} \ \  \ \ \ \
\subfigure[$c$ vs $d$ cross section for a fixed $d$.
  The critical exponent is $0.5 \pm 0.0001$ near phase transition point  $E=1$ .]  
  {\includegraphics[width=6cm]{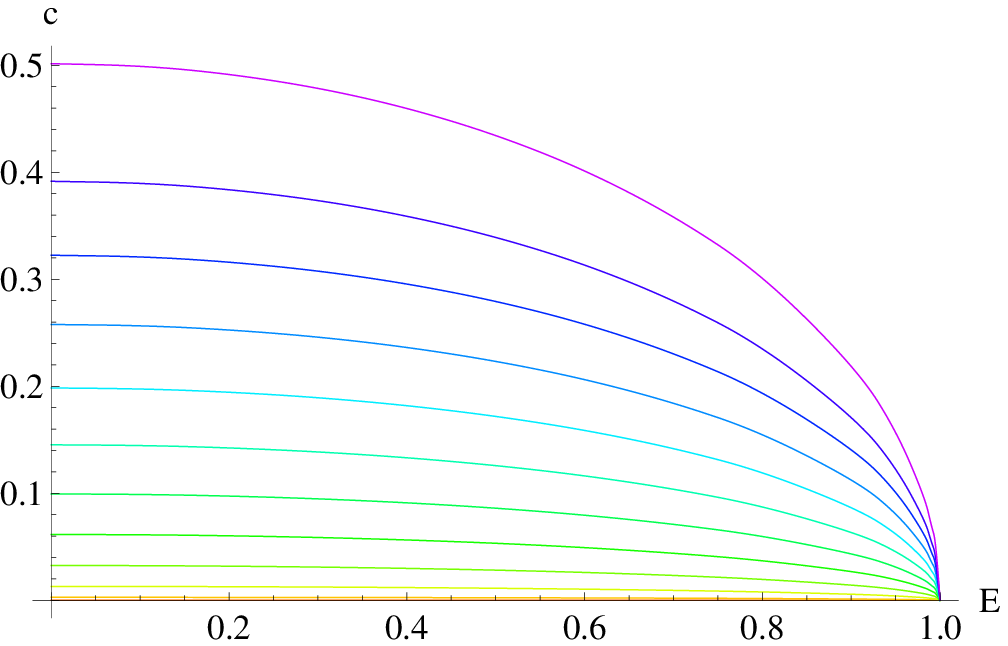}}  
\subfigure[$c$ vs $d$ cross-section near phase transition $d=\sqrt{7}$ for a fixed $E$. $E=0.1,0.2 , \cdots, 
   0.9$ from orange to purple. BKT scalings are shown.   
     the BKT transition point $d_c = \sqrt{7}$ is independent of $E$ 
     as shown with the IR analysis.]
  {\includegraphics[width=7cm]{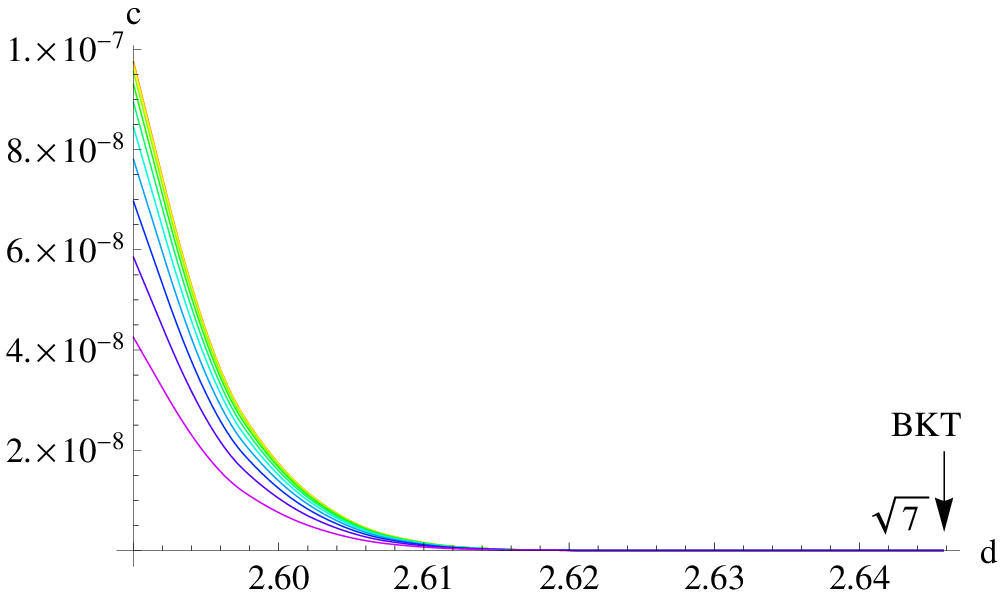}} \ \ \ \ \ \
  \subfigure[Numerical confirmation of BKT scaling. $c \sim e^{-\pi\sqrt{\frac{1+d^2}{d_c^2-d^2}} }$.
  The slopes are $1 \pm 0.001$.]
  {\includegraphics[width=6cm]{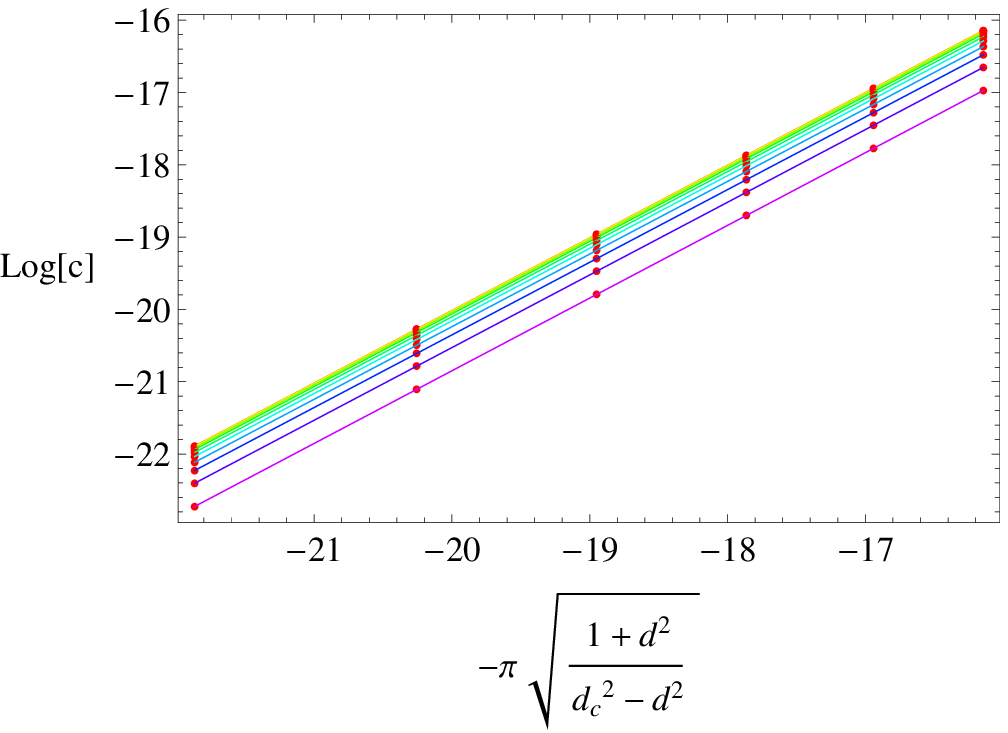}}
  \caption{ The analysis of the BKT and mean field phase transition for the $D3/D5$ system. $B=1$ and $T=0$.}\label{cvsd}
\end{figure}

Having tamed the numerics we are able to study the phase transition from the chirally symmetric phase to the chiral symmetry breaking phase and this is plotted in Fig \ref{Ed}.
For the $D5$ case the analysis of the condensate as a function of $d$ and $E$ is shown in Fig \ref{cvsd}. 
The phase transition as $d$ is increased is independent of the size of the electric field (for $\tE<1$) and is of BKT type. As $\tE$ increases for $d<d_c=\sqrt{7}$ the phase transition is of mean field type and again its position is independent of the value of $d$. This should be contrasted with the $D7$ system where the second order phase transition takes a non-trivial profile in the $d-E$ plane. The $D7$-brane analysis is relatively simple and so is not presented here.



The line of BKT transitions in the $D3/D5$ system and its $\tE$-independence can be understood from the IR dynamics. After scaling all quantities by $\tB$ we have the embedding Lagrangian, 
 at small $\rho$,   
\begin{equation}
  \call_{LT} \sim \sqrt{1-\tE^2} \sqrt{d^2 
  + \rho^{4}{(\rho^2 + L^2)^2}}
  \sqrt{1+L'^2} \,.
\end{equation}

Note that the $\tE$ part is factored out from the Legendre transformed action so that the dynamics is identical to the $\tE=0$ case. The slipping mode for a small fluctuation of the embedding $L$  
in the IR $\rho \ll (1, d)$ behaves as a scalar in AdS$_2$ space 
with mass $-\frac{2}{1+d^2}$.   
The Breitenlohner-Freedman (BF) bound for a scalar in AdS$_2$ is $-1/4$, 
so below $d = \sqrt{7}$ the BF bound is violated and the 
phase transition to a broken symmetry phase occurs. 
Note that we get a good fit for the whole of the region close to the phase transition with $c\sim 2.6 e^{\frac{-\pi\sqrt{1+d^2}}{\sqrt{7-d^2}}}\sqrt{1-\tE}$. The exponent comes from the analytic calculation given in \cite{Jensen:2010ga}.

\subsection{$T$-$d$-$E$ phase diagram} \label{sec3D}
\begin{figure}[]
\centering
\subfigure[D3/D5]
  {\includegraphics[width=7cm]{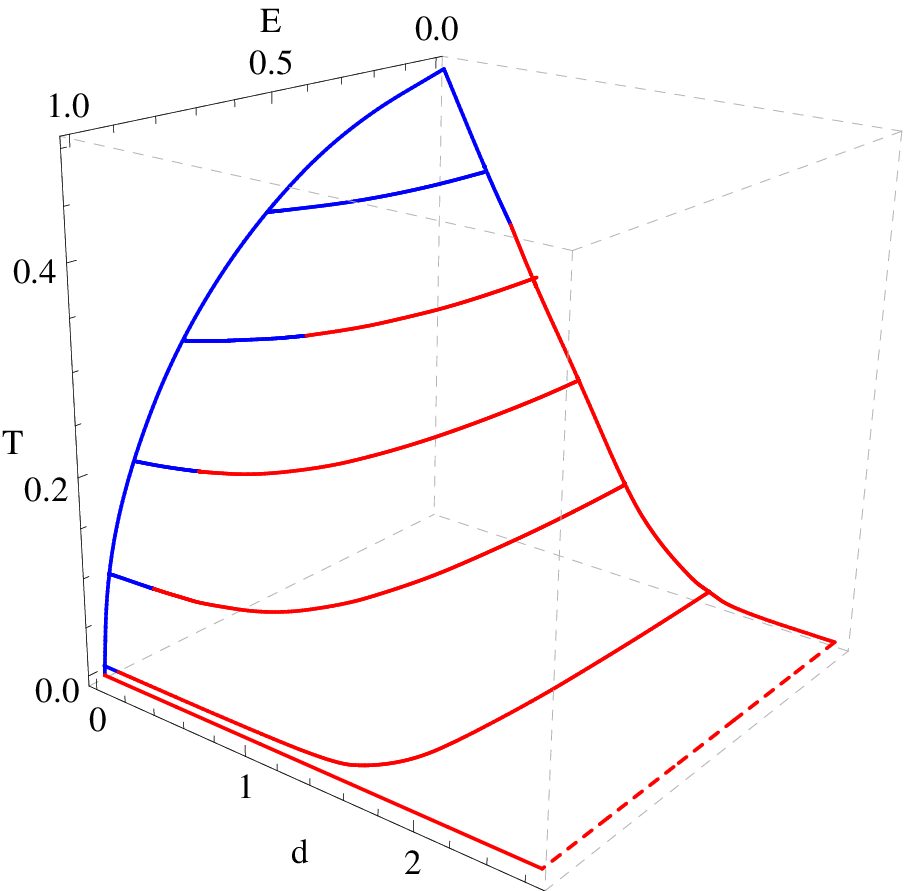}}
\subfigure[D3/D7]
  {\includegraphics[width=7cm]{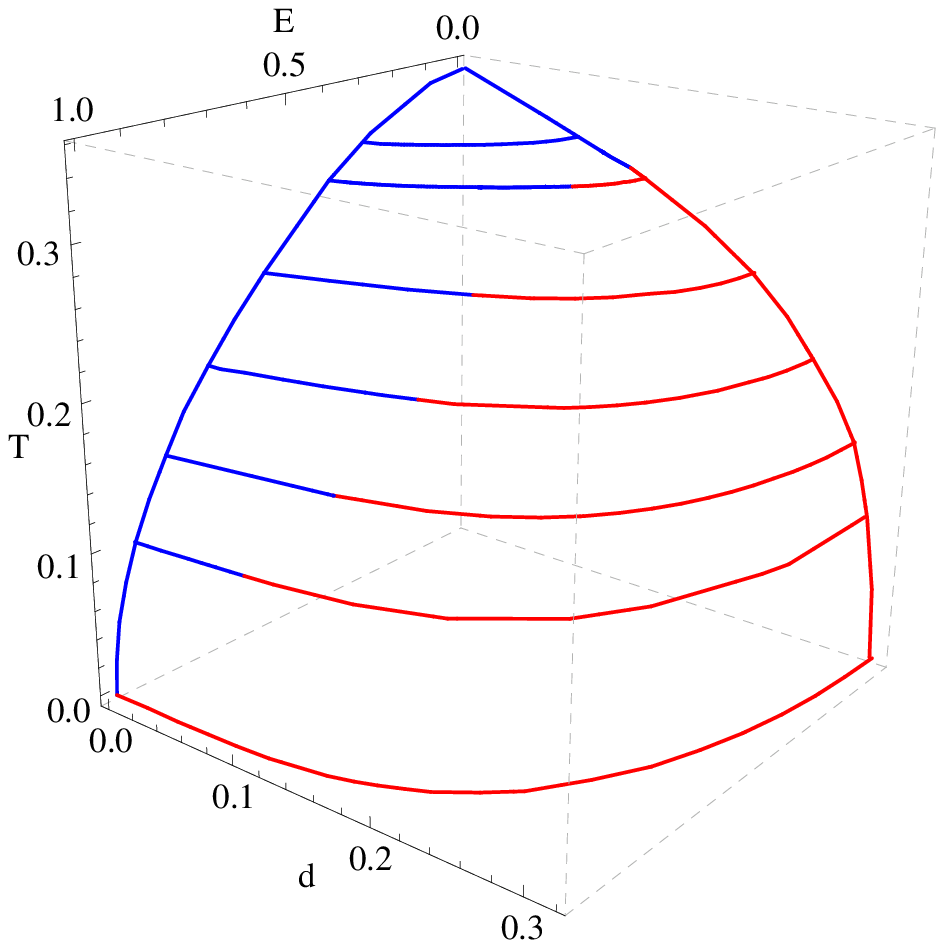}}  
  \caption{ 3D phase diagram of D3/D7 and D3/D5 system at $B=1$ as
  a function of $T$,$d$, and $E$. } \label{3D}
\end{figure}

Finally, in Fig \ref{3D} we combine the above information with the phase structure outside the two dimensional slices we have analyzed above. The critical point linking the first and second order phase transitions give a line on the surface of the phase transition face. This links surfaces of second order and first order phase transitions. The line of BKT phase transitions is an isolated region in the $D3/D5$ diagram and the rest of the diagram is either of mean-field second order type or of first order type. Fig \ref{3D} summarizes our full analysis. 

\section{Summary}

We have studied the D3/D7 and D3/D5 systems in the presence of mutually perpendicular electric
and magnetic fields to deduce the temperature and density phase structure of the dual gauge theories.
Those theories are the 3+1 dimensional ${\cal N}=4$ super Yang Mills theory  with quark hypermultiplets
introduced in all field theory directions and on a 2+1 dimensional defect respectively. The full results are displayed, for constant
magnetic field, in Fig \ref{3D}. There are two phases. At large $d,T,E$ the theory is in
a chirally symmetric phase, while at low $d,T,E$ the
phase displays chiral symmetry breaking through a non-zero quark condensate. 
In the gravity
description these transitions are associated with embeddings of the probe D7 or
D5-brane that either lie flat (the symmetric phase) or  are given by non-trivial profiles in the holographic
directions breaking the symmetry.

At zero temperature the phase transition is continuous for both D3/D7 and D3/D5 systems. 
However, the D3/D5 system has a special feature that there is a holographic BKT transition for low $E$ field values. In this regime the deep IR is
dominated by both the magnetic field and density regardless of $E$. 
When the electric field becomes sufficiently large to trigger the transition on its own the transition becomes second order and mean field. 
At intermediate temperature,  at low density the
transition is first order whilst at high density it is mean-field second order. The critical lines between these regimes can be inferred from the Fig \ref{3D}.  
At high temperature the transition becomes all first order. 

If we were to include finite mass, the structure may have even more subtleties within. The more features we can find in our phase diagrams, the more chance there is we can link such basic holographic models with real world systems. We leave the extensions of this phase space for future work.

\acknowledgments

NE is grateful to the support of an STFC rolling grant. KK is grateful for support by the University of Southampton.
J.S. is supported by the EU with a Marie Curie Fellowship and is grateful to the Max Planck Institute for physics in Munich. 
We thank Andy O'Bannon, Veselin Filev, and Da-Wei Pang for discussions.

\providecommand{\href}[2]{#2}\begingroup\raggedright

\endgroup

\end{document}